\documentclass[copyright,creativecommons,noderivs,noncommercial]{eptcs}
\providecommand{\event}{Gabriel Ciobanu (Ed.): Membrane Computing and\\ Biologically Inspired Process Calculi 2009} 
\providecommand{\volume}{15}   
\providecommand{\anno}{2009}   
\providecommand{\firstpage}{1} 
\providecommand{\eid}{1}       

\usepackage{breakurl}

\usepackage{verbatim}
\usepackage{graphicx}
\usepackage{amsmath, amssymb}
\usepackage{longtable}
\usepackage{array}
\usepackage{txfonts}
\usepackage{empheq}
\usepackage{setspace}

\usepackage{pepa,tree}
\usepackage{subfigure}

\newcommand{\Real}{{\mathbb R}}

\newcommand{\reactant}{\mbox{$\downarrow$}}
\newcommand{\product}{\mbox{$\uparrow$}}

\newcommand{\modifier}{\odot}
\newcommand{\activator}{\oplus}
\newcommand{\inhibitor}{\ominus}
\newtheorem{definition}{\textbf{Definition}}

\title{A compartmental model of the cAMP/PKA/MAPK pathway in Bio-PEPA}
\author{Federica Ciocchetta
\institute{The Microsoft Research - University of Trento Centre\\for Computational and Systems Biology,  Trento, Italy}
\email{ciocchetta@cosbi.eu}
\and
Adam Duguid
\institute{University of Edinburgh, UK}
\email{a.j.duguid@sms.ed.ac.uk}
\and
Maria Luisa Guerriero
\institute{University of Edinburgh, UK}
\email{mguerrie@inf.ed.ac.uk}
}

\def\titlerunning{A compartmental model of the cAMP/PKA/MAPK pathway in Bio-PEPA}
\def\authorrunning{F. Ciocchetta \& A. Duguid \& M.L. Guerriero}

\begin{document}
\maketitle

\begin{abstract}
The vast majority of biochemical systems involve the exchange of
information between different compartments, either in the form of transportation or via the intervention of membrane proteins which are able to transmit stimuli between bordering compartments.

The correct quantitative handling of compartments is, therefore,
extremely important when modelling real biochemical systems. The
Bio-PEPA process algebra is equipped with the capability of
explicitly defining quantitative information such as compartment
volumes and membrane surface areas. Furthermore, the recent
development of the Bio-PEPA Eclipse Plug-in allows us to perform a
correct stochastic simulation of multi-compartmental models.

Here we present a Bio-PEPA compartmental model of the cAMP/PKA/MAPK pathway.
We analyse the system using the Bio-PEPA Eclipse Plug-in and we show
the correctness of our model by comparison with an existing ODE
model. Furthermore, we perform computational experiments in
order to investigate certain properties of the pathway. Specifically,
we focus on the system response to the inhibition and strengthening
of feedback loops and to the variation in the activity of key
pathway reactions and we observe how these modifications affect the
behaviour of the pathway. These experiments are  useful to
understand the control and regulatory mechanisms of the system.
\end{abstract}

\section{Introduction}
\label{sec:introduction}

Compartments are widely present in biological systems and play a major
role in a lot of biological processes~\cite{albertsEtAl02}; they are enclosed by membranes, which isolate biological species from the external environment.  Biological species can be either located inside a compartment or reside on a membrane; in the latter case they can be attached to a side of the membrane~(\emph{receptors}, for instance) or span the entire membrane~(i.e.\ \emph{transmembrane proteins}).  Species located in the same compartment can interact with each other through biochemical reactions; species can also move from one compartment to another via diffusion or through membrane channels, thus allowing the passing of information between adjacent compartments.
Alternatively, the transmission of signals between compartments can involve receptors, which respond to the
input of signalling molecules on one side of the membrane by triggering a cascade of events on the other side.

Due to the fundamental importance of compartments, their correct
handling is often essential when studying biological systems, and an
increasing number of modelling languages support their explicit
definition~\cite{paun-rozenberg02,cardelli04,regevEtAl04,priami-quaglia05a,Cavaliere06,versari08}.
These languages differ from each other in the kind of abstraction,
the assumptions made~(for instance, static or dynamic compartments)
and the features they are designed to represent. Most of them are
supported by analysis
tools~\cite{BAMHP,frameworkHP,betawbHP,cytosimHP}.

In this work we consider Bio-PEPA with locations as our modelling language~\cite{Ciocchetta-Guerriero09}. Bio-PEPA locations abstract both biological compartments and membranes: they are defined by \emph{names}, enriched with additional information allowing the modeller to express their kind~(i.e.\ compartment or membrane), their size, and their position with respect to the other locations of the system.

Bio-PEPA offers a high level of abstraction, similar to that used to describe biochemical networks in the literature and in biological databases~(see, for instance, models in the \emph{BioModels} database~\cite{lenovereEtAl06}). Only \emph{static locations} can be represented~(i.e.\ compartments cannot merge, split, or undergo any structural change); nevertheless, their size can vary with respect to time.

The main aim of this work is to demonstrate the power of Bio-PEPA with locations as a modelling language for multi-compartmental biochemical networks. Thus, we consider a moderately sized signalling pathway describing the cAMP/PKA/MAPK activity~\cite{NevesEtAl08} with the explicit definition of  compartments and membranes and we translate it into Bio-PEPA. In this pathway the spatial aspects are particularly relevant, since it involves interaction between molecular species residing in different biological compartments with different volume sizes.

The analysis of the model is performed using the Bio-PEPA Eclipse Plug-in~\cite{ciocchettaEtAl09_tools}, a novel simulation tool for the construction and analysis of Bio-PEPA models.  The Bio-PEPA Eclipse Plug-in offers methods for numerical integration of Ordinary Differential Equations~(ODEs) and stochastic simulation. We present here the latest version of the tool, which supports the definition of multi-compartmental models defined using the Bio-PEPA extension with locations and is able to handle multiple compartments with different sizes in the appropriate way.

Using the Bio-PEPA Eclipse Plug-in, we validate the model against the expected results in the literature and we perform in-silico experiments in order to investigate some properties of the pathway. Specifically, we study how the behaviour of the pathway is affected by varying the enzymatic activity of three key molecular species and by the removal of feedback loops with an important regulatory effect over particular species. These experiments are useful for a better understanding of the regulatory and control mechanisms of the system.

The structure of the paper is as follows. Section~\ref{sec:biopepa}
reports a brief introduction of the Bio-PEPA language, while
Section~\ref{sec:tool} is devoted to the description of the tool. In
Section~\ref{sec:biomodel}  we present a model of the cAMP/PKA/MAPK
signalling pathway, and in Section~\ref{sec:analysis} we show the
validation of the model and some analysis results.
Section~\ref{sec:relatedworks} concerns some related work, whereas
the last section reports some conclusive remarks. The full Bio-PEPA
model of the pathway is reported in~Appendix~\ref{sec:appendix}.

\section{Bio-PEPA with locations}
\label{sec:biopepa}

\newcommand{\transport}{\mbox{$\rightarrow$}}
\newcommand{\rtransport}{\mbox{$\leftrightarrow$}}
\newcommand{\eqdot}{=\!\!\!\!\!\cdot\,\,\,\,\,}
\newcommand{\eqdotbox}{\stackrel{\doteq}{\boxdot}}
\def\Ss{\mbox{\large $\rhd\!\!\!\!\lhd$}}
\newcommand{\syncs}[1]{\raisebox{-0.9ex}{$\:\stackrel{\Ss}{\scriptscriptstyle #1}\,$}}
\newcommand{\syncbis}[1]{\raisebox{-1.0ex}{$\;\stackrel{\Ss}{\scriptstyle #1}\,$}}
In this section we describe the main features of the extension of Bio-PEPA~\cite{ciocchetta-hillston09} with locations~\cite{Ciocchetta-Guerriero09},
the modelling language used in this work.

A Bio-PEPA system representing a biochemical network consists of  a \textit{context}, defining information such as locations, kinetic rates, parameters and auxiliary information, a
set of \textit{species components} and a \textit{model component}.

Locations represent biological compartments and membranes. They are
described by \textit{names}, enriched with additional information
allowing the modeller to express their position with respect to the
other locations of the system, their kind~(i.e.\ compartment or
membrane), and their size~(i.e.\ volume or surface area). The
structure of the biological system is modelled as a \textit{static
hierarchy}, represented as a tree whose nodes represent locations;
each node has one child for each of its sub-locations. The
\emph{location tree} allows us to keep track of the relative
positions of locations and must be associated with the location
definition.

The locations are defined as follows.

\begin{definition}
\label{Def:location}
Each location is described by ``$L: \; size\;  unit, \; kind$'', where
$L$ is the~(unique) location name, ``$size$'' expresses the size and
can be either a positive real number or a more complex mathematical
expression depending on time $t$; the~(optional) ``$unit$'' denotes
the unit of measure associated with the location size, and
``$kind$'' $\in\{\mathbf{M}, \mathbf{C}\}$ expresses if it is a
membrane or a compartment, respectively. The set of locations is
denoted by $\mathcal{L}$.
\end{definition}

Functional rates are used to describe kinetic laws associated with a reaction. They have the form ``$f\_{\alpha} = \textit{expr}$'', where $\alpha$ is the action type abstracting the reaction and \textit{expr} is a mathematical expression describing the law. The kinetic parameters used in the mathematical expressions are defined by parameters in Bio-PEPA.

Species components describe the behaviour of the individual biological species, whereas the model component defines the interactions between the various species and the amount of them present in the system.
The syntax of the Bio-PEPA components is defined as:
$$
S ::= (\alpha, \kappa) \mbox{ \texttt{op} } S \ \mid \ S + S \ \mid \ C \ \mid \ S@location \quad  \mbox{with }\texttt{op} = \reactant \mid \product \mid
\activator \mid \inhibitor \mid \modifier
\quad \quad \quad P::= S(x) \mid P \syncs{\mathcal{I}}  P
$$
\noindent where $S$ is the
\emph{species component} and $P$ is the \emph{model component}. In
the prefix term $(\alpha,\kappa) \mbox{ \texttt{op} } S$, $\kappa$
is the \emph{stoichiometry coefficient} of species $S$ in reaction
$\alpha$, and the \emph{prefix combinator} ``\texttt{op}'' represents the
role of $S$ in the reaction. Specifically, $\reactant$ indicates a
\emph{reactant}, $\product$ a \emph{product}, $\activator$ an
\emph{activator}, $\inhibitor$ an \emph{inhibitor}, and $\modifier$ a
generic \emph{modifier}.
We can use ``$\alpha \mbox{ \texttt{op} }S$'' as an abbreviation for
``$(\alpha, 1) \mbox{ \texttt{op} } S$''.
The operator ``$+$'' expresses the choice
between possible actions, and the constant $C$ is defined by an
equation $C \rmdef S$.  The term $S @location$
indicates that the species represented by the process $S$ is the location $location$. The parameter $x \in \Real^+$ in $S(x)$
represents the number of molecules, which can be abstracted to a level~\cite{ciocchettaEtAl08FBTC} for analysis as a CTMC. Finally, the process $P
\syncs{\mathcal{I}} Q$ denotes the cooperation between components:
the set $\mathcal{I}$ determines those activities on which the
operands are forced to synchronise. The notation $\syncbis{*}$ means that the species components are obliged to synchronize on the common action types.

Note that the same species in different locations is represented by distinct species components. Therefore, in order to avoid the possible duplication of actions in the model, we propose a notation to represent a species in
multiple locations in a compact way. Given a species $S$ which can be in $L_1, \dots, L_n$, one single
component $S$ is defined. This is simply a shorthand
for a set of definitions $S @ L_1, \dots, S @ L_n$, each of which
contains only those actions which can occur in the respective
location. A reaction $\alpha_1$ occurring only in
$L_i$ is defined as $(\alpha_1, \kappa) \mbox{ \texttt{op} } S @
L_i$, while $\alpha_2$ occurring in all locations $L_1, \dots , L_n$
is defined as $(\alpha_2, \kappa) \mbox{ \texttt{op} } S$.

No constraint is imposed on the location of interacting species. This implies that it is possible to describe reactions involving species located in non-adjacent locations. The reason behind this choice is that, even though interactions are generally local~(i.e.~involve molecules in the same or adjacent locations), imposing constraints on the relative position of species locations could prevent the modellers from defining some reactions~(e.g.~a reaction abstracting a sequence of reaction steps involving various locations).

The formal definition of the Bio-PEPA system is the following.

\begin{definition}
\label{def:biopepa} A Bio-PEPA system $\mathcal{P}$ is a 6-tuple
$\langle \,\mathcal{L},\mathcal{N},\mathcal{K}, \mathcal{F}_R,Comp,P
\rangle$, where: $\mathcal{L}$ is the set of locations,
$\mathcal{N}$ is the set of~(auxiliary) information for the species,
$\mathcal{K}$ is the set of parameters, $\mathcal{F}_R$ is the set
of functional rates, $Comp$ is the set of species components and $P$
is the model component.
\end{definition}

A main feature of Bio-PEPA is that it can be seen as a formal,
intermediate representation of biochemical systems, on which
different kinds of analysis can be carried out, through the defined
mappings into contin\-uous-deterministic and discrete-stochastic
modelling languages and CTMC with levels.

Bio-PEPA is given an operational semantics, which is used for the
derivation of the stochastic labelled transition system and the
associated CTMC with levels. We do not report here the details on
the operational semantics since they are not needed for the kinds of
analysis on which we focus in this work~(continuous-deterministic
and discrete-stochastic). For details,
see~\cite{Ciocchetta-Guerriero09,ciocchetta-hillston09}.

Finally, it is important to remember that Bio-PEPA~(and, in general, process algebras) offers a
\emph{compositional approach} for the definition of models: it is possible to build the full model starting from the definition of its subcomponents. This approach is particularly useful for biological systems, as they are generally large and complex. For these systems, a compositional definition of the model is desirable since it makes the construction of the full model easier and it helps in understanding  the relationships between its various subcomponents and the structure of the network. The compositional approach has been used in the definition of the Bio-PEPA model for the cAMP/PKA/MAPK pathway.
The compositional rules for Bio-PEPA models are standard and not reported here.

\newcommand{\sub}[1]{\ensuremath{{}_{\textrm{#1}}}}
\newcommand{\super}[1]{\ensuremath{{}^{\textrm{#1}}}}
\section{The Bio-PEPA Eclipse Plug-in}\label{sec:tool}
The Bio-PEPA Eclipse Plug-in~\cite{biopepa_site} is one of two tools developed at Edinburgh University for the Bio-PEPA language. Built on top of the Eclipse platform~\cite{Eclipse}, the plug-in offers an operating system agnostic editor and suite of time-series solvers for the construction and analysis of Bio-PEPA models. It provides the modeller with a rich modelling environment for Bio-PEPA, from the initial creation of the model through to the analysis and inspection of the results.

\begin{figure}[htb]
\begin{center}
\includegraphics[width=\textwidth]{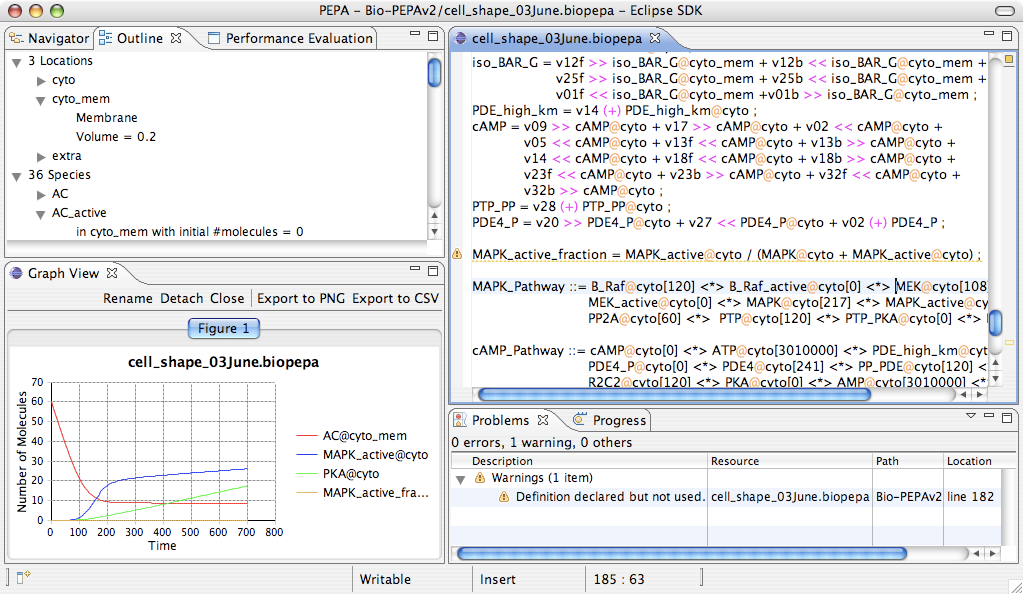}
  \caption{Screen shot of the Bio-PEPA Eclipse Plug-in. The editor can be seen in the top left corner, with the outline view in the top right. The problems tab~(bottom right) displays warnings and errors associated with the open models and the ability to plot results can be seen in the bottom left.}
\label{fig:screenshot}
\end{center}
\end{figure}

The features currently supported in the plug-in can be roughly grouped into those assisting the modeller in the writing of their models and features for the analysis of the model. Examples of the former include an editor~(with Bio-PEPA-aware syntax highlighting) and static analysis of the model for detecting various syntactic errors. For the latter, the plug-in is capable of generating and displaying time-series results, from ODE numerical integration or stochastic simulation algorithms, with the option to export these results in both graphical and textual formats. Many of these features can be seen in Figure~\ref{fig:screenshot}.

As with all previous releases of the plug-in, the language is augmented with keywords for particular definitions. These keywords increase the readability of the model, clearly signposting those definitions pertaining to locations, species and kinetic laws. These indicators are then emphasised further by the syntax highlighting built into the developing environment. Together they assist the modeller in recognising the purpose of various definitions found within any given model. As of version 0.1.0, the Eclipse Plug-in will contain initial support for multi-compartmental models as defined by the Bio-PEPA extension with locations. The following description details both this support along with several refinements to the implemented language. For the sake of brevity the description does not cover the entire language, instead concentrating on those required to model the system of interest.

Units of measures are currently unsupported, all values must be defined with respect to the same generic unit, with compartments referring to volumes~(unit\super{3}) and membranes as surface area~(unit\super{2}). Finally, the initial values for species are assumed to be in terms of molecular counts. With the introduction of locations with different sizes, concentrations become an inappropriate form of measurement and must be translated into molecular counts.

\subsection{Supported Language}
Bio-PEPA does not impose a strict ordering on the definitions of context and species components, but these must all precede the model component. This final statement provides the structure of the system as well as the initial conditions. Since the modeller is relatively unconstrained with respect to the ordering we shall present the syntax for different definitions in the order that they are parsed within the plug-in. The first is the location definition, an example of which can be seen below:
\begin{eqnarray*}
\mathrm{location} \; \mathit{cyto} \; \mathrm{in} \; \mathit{cyto\_mem} & :  &
\mathrm{size} = 1, \; \; \mathrm{kind} = \mathit{compartment};\\
\mathrm{location} \; \mathit{cyto\_mem} & : & \mathrm{size} = 0.2, \; \; \mathrm{kind} = \mathit{membrane};
\end{eqnarray*}
The required keyword \texttt{location} identifies this as a location statement and the optional keyword \texttt{in} is used for defining the position of the given location in the location tree. In the example above we have defined a compartment $cyto$ to exist within a second location known as $cyto\_mem$. Allowable parameters for these locations include the size and kind, both of which can be seen. For reasons of legibility acceptable kinds are \texttt{compartment} and \texttt{membrane} rather than $\mathbf{M}$ and $\mathbf{C}$. Based on the kinds in the example and the spatial hierarchy we can clearly see that $cyto\_mem$ is the membrane to the $cyto$ compartment.

The next required definition in a Bio-PEPA model describes the kinetic rates of the reactions, with an example below:
\begin{eqnarray*}
\mathrm{kineticLawOf} \; \mbox{ \textit{v1f} } & : &
kf\_activate\_Gs * iso\_BAR\_G@cyto\_mem;
\end{eqnarray*}
\noindent where $kf\_activate\_Gs $ is the constant rate of the reaction \textit{v1f} and \textit{iso\_BAR\_G@cyto\_mem} is the reactant of the reaction~(the species amount is given in number of molecules).
Like location definitions, this definition is easily identifiable by the required keyword \texttt{kineticLawOf}, followed by its label for reference within the species component definitions. The defined rate can either be a custom rate, as seen in the example, or make use of predefined laws such as mass-action or Michaelis-Menten. For example, if we know that the only reactant in the reaction associated with \textit{v1f} is $iso\_BAR\_G@cyto\_mem$, then this custom rate describes mass-action. This could then be re-written as:
\begin{eqnarray*}
\mathrm{kineticLawOf} \; \mbox{ \textit{v1f} }  & :  &
\mathrm{fMA}(kf\_activate\_Gs);
\end{eqnarray*}
\noindent where $\mathrm{fMA}(r)$ stands for mass-action kinetic law with constant parameter $r$.
The inclusion of predefined laws offers two clear benefits to the modeller. Mistakes in inputting rates can be reduced by removing the need to input the laws manually. Secondly, the use of predefined kinetic laws allows for additional static analysis of the model. If a reaction is defined as using Michaelis-Menten~($\mathrm{fMM}(v_{M}, k_{M})$, where $v_{M},\, k_{M}$ are the two constants in Michaelis-Menten kinetic law\footnote{Given an enzymatic reaction $S + E \rightarrow P + E$, where $S$ is the substrate, $P$ is the product and $E$ is the enzyme, the Michaelis-Menten kinetic law is $v_M \cdot E \cdot S /(K_M +S)$, where $v_M$ is the maximum rate constant and $K_M$ is the Michaelis constant.}), then the reaction must have a single species acting as a reactant or substrate, another species performing the role of the catalyst or enzyme and a third as the product of the reaction. If the list of species involved in this reaction differs from those expected for the kinetic law stated the tool will flag this as an error for the modeller.

For the species component definitions, acceptable symbols for the different possible behaviours  are required. Below is the list of Bio-PEPA operators along with their ASCII equivalent for creating Bio-PEPA models.

\begin{center}
\begin{tabular}{l | c | c}
Behaviour & Bio-PEPA symbol & ASCII representation\\
\hline
reactant & $\reactant$ & $<<$\\
product & $\product$ & $>>$\\
activator & $\activator$ & $(+)$\\
inhibitor & $\inhibitor$ & $(-)$\\
modifier & $\modifier$ & $(.)$\\
uni-directional transportation & $\transport$ & $->$\\
bi-directional transportation & $\rtransport$ & $<->$\\
cooperation & $\sync{\mathcal{I}}$ & $<I>$
\end{tabular}
\end{center}

These operators are used within the species component definitions, an example of which can be seen below.
\begin{eqnarray*}
PDE4 & = & v27 >> PDE4@cyto \; + \; v20 << PDE4@cyto \; + \; v5 \: (+) \: PDE4@cyto;\\
\end{eqnarray*}
\noindent  The species PDE4 is involved in three interactions with different roles: it is a product in the reaction \textit{v27}, a reactant in \textit{v20} and an enzyme in \textit{v5}. In the example above all three reactions are only possible when the species PDE4 is in the location \textit{cyto}.
From this example we can see shorthand notations for when the stoichiometric coefficient is one. As with all defined shorthands, their purpose is to keep the definitions free of superfluous details. This is not to suggest that stoichiometric information in general is unnecessary, but that through the use of this shorthand we can emphasise the less common instances~(a stoichiometry of one being the majority of observed values) where this is not the case.

The final definition in a Bio-PEPA system is that of the model component~(see Section~\ref{sec:biopepa}).  The model component describes the initial amount of each species in the system and the possible interactions involving these species. The plug-in however offers true compositionality. It does this by first augmenting the model component syntax to accept names and also allowing definitions written in the syntax of the model component to be assigned names. These additional labelled compositional definitions can then be referred to in either the model component or other labelled compositional definitions. This allows the modeller to describe a complex system in terms of smaller pathways, or indeed any logical grouping that they wish to enforce.

\begin{eqnarray*}
MAPK\_Pathway & ::= & B\_Raf@cyto[120] <*> B\_Raf\_active@cyto[0] <*>  \dots\\
&& <*> MAPK@cyto[217] <*> MAPK\_active@cyto[0]\\
&& <*> \dots <*> PTP\_PP@cyto[60];
\end{eqnarray*}

Here we have a labelled composition definition for the MAPK pathway. It should be noted that labelled compositional definitions use a different assignment operator to other definitions, using $::=$ instead of $=$. As stated earlier, and as can be seen in the example above, labelled compositions also use the definition terminator. This leaves the model component as the final definition to present, which an example of can be seen below:

\begin{displaymath}
G\_Pathway <v09, v17> cAMP\_Pathway <v08, v15> MAPK\_Pathway
\end{displaymath}

The benefits gained through compositional definitions are highlighted in this model component definition. Not only does the model component make it clear that the system is compromised of three pathways, but can also serve to highlight important reactions. We can see that interaction between the cAMP and MAPK pathways is limited to reactions $v08$ and $v15$. If the MAPK pathway is reliant on the cAMP pathway, rather than any interaction being redundancy in the system, then the rates for these reactions are likely candidates for perturbation. This of course is not guaranteed, merely that the cooperation sets can be used to highlight limited interactions between groups of species. The compositional approach also makes it very easy to enable or disable entire pathways.

Once the model has been built, the plug-in parses the model and performs static analysis to detect a variety of modelling errors, an example of which was given earlier in regards to kinetic rate laws. Once the model is free of these errors the modeller has access to ODE solvers in the form of a Runge-Kutta implicit-explicit solver and a Dormand-Prince adaptive step-size solver~\cite{stoer93}, these being made available through the odeToJava library. The plug-in also supports analysis through stochastic simulators, with implementations of Gillespie's Stochastic Simulation Algorithm~(SSA)~\cite{gillespie77}, Gibson-Bruck~\cite{gibson-bruck00} and the Tau-Leap algorithms~\cite{gillespie-petzold03} available. The ability to use different types of solvers has allowed us to discover errors in published modelling studies in computational biology~\cite{calderEtAl06}.

\section{The cAMP/PKA/MAPK pathway}
\label{sec:biomodel}

The cAMP/PKA/MAPK network is a ubiquitous signalling pathway. In
this work we consider the system as modelled in~\cite{NevesEtAl08}
and in particular the ODE model described by the authors and
available in SBML format from~\cite{sbml_NevesEtAl08}.

The cAMP-dependent protein kinase~(PKA) is responsible for
performing several functions in the cell, such as regulation of
glycogen, sugar, and lipid metabolism. Among the downstream targets
of PKA there are mitogen-activated protein kinases~(MAPKs), which
are essential in regulating various cellular activities, including
gene expression, mitosis, differentiation, and cell
survival/apoptosis.

The major role of this pathway in various cellular processes,
together with the availability of quantitative data such as the
sizes of the involved compartments, make it an interesting case
study for our purposes.

As is common in signalling pathways, spatial aspects are particularly
relevant in this system, since it involves interaction between
molecular species residing in different biological compartments. We
consider molecules lying inside the cell~(i.e.\ in the cytoplasm),
and outside it~(i.e.\ in the extra-cellular environment),
specifically the receptor agonist which triggers the signalling
cascade. These two compartments are separated by a membrane~(i.e.\
the cytoplasmic membrane) where receptors and specific
membrane-proteins are located.

The compartments have different sizes and this exercises a direct
impact on the quantitative behaviour of the system. As in the
original ODE model~\cite{NevesEtAl08,sbml_NevesEtAl08}, we consider
the cell to be a sphere with a surface-to-volume ratio equal to 0.2.
The kinetic parameters we use are taken from the original paper,
which have been derived from the biochemical literature.

Before presenting the Bio-PEPA model in the following
Section~\ref{sec:biomodel_biopepa}, we briefly describe the
biological system. Figure~\ref{fig:pathway} shows a schematic
representation of the pathway, and the full model is reported in
Appendix~\ref{sec:appendix}.

\begin{figure}[htbp]
\centering
  \includegraphics[width=0.8\textwidth]{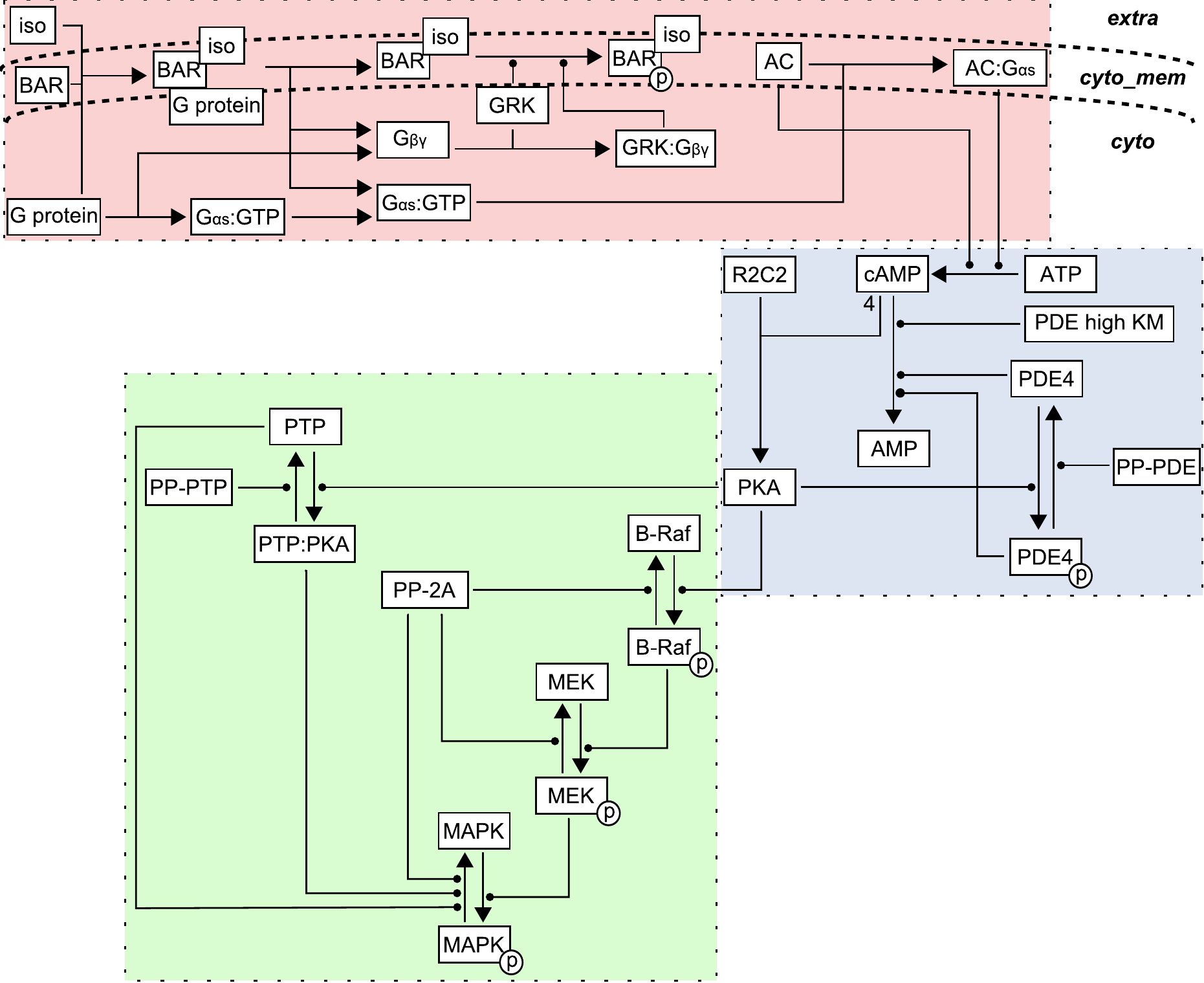}\\
  \caption{Schematic representation of the pathway. The coloured boxes represent a decomposition of the pathway into three subnetworks:
  the G-protein mediated activation of AC, the activation of PKA, and the MAPK signalling cascade.
  Arrows represent reactions, and lines terminating with circles represent enzymes participating in reactions.}\label{fig:pathway}
\end{figure}

The cAMP/PKA/MAPK network we consider is in neurons, and its
biological action is initiated by isoproterenol~(ISO) binding to
specific cell-surface receptors called $\beta$-adrenergic receptors
(BARs), which are a class of G-protein coupled receptors~(GPCRs).

GPCRs reside in the cytoplasmic membrane and respond to a wide range
of extracellular stimuli, in our case to the synthetic drug
isoproterenol. When GPCRs bind an extracellular ligand, they
interact with G-proteins and propagate the signal across the
membrane into the cytoplasm.

G-proteins are heterotrimeric complexes bound to the inside surface
of the cell membrane, and consist of an active subunit $G_{\alpha}$
and two tightly associated $G_{\beta}$ and $G_{\gamma}$ subunits.
$G_{\alpha}$ subunits alternate between an inactive guanosine
diphosphate~(GDP) and an active guanosine triphosphate~(GTP) bound
state. The transition from the inactive to the active state of
$G_{\alpha}$ and its simultaneous dissociation from the
$G_{\gamma\beta}$ subunit and the GPCR are stimulated by the binding
of GPCRs with the ligand.

$G_s$ subunits are a subclass of $G_{\alpha}$ subunits, and they
stimulate the production of cyclic-AMP~(cAMP) from ATP, by
activating~(by binding to it) the membrane-associated enzyme
adenylate cyclase~(AC) which, in turn, produces cAMP.

cAMP is a key signalling molecule which acts as a second messenger,
activating the cAMP-dependent protein kinase A~(PKA). PKA is a
tetrameric holoenzyme, consisting of two regulatory and two
catalytic subunits~($R_2C_2$). Under low levels of cAMP, PKA is
inactive; when the concentration of cAMP rises, cAMP molecules bind
to each of the two binding sites on the two regulatory subunits,
which leads to the dissociation and activation of the catalytic
subunits~(i.e.~four cAMP molecules are needed to activate one PKA
molecule). Upon activation, the catalytic subunits can phosphorylate
a number of downstream targets.

PKA is an enzyme which participates in several metabolic and
signalling pathways, leading to a wide range of biological
responses. Here we focus on its ability to initiate the pathway
leading the activation of the Mitogen-Activated Protein
Kinase~(MAPK).

The downstream pathway in our network represents an instance of the
common MAPK signalling chain. It starts with the phosphorylation of
the MAPKKK B-Raf, which phosphorylates~(i.e.~activates) the MAPKK
MEK, which in turn phosphorylates MAPK.

The amounts of PKA and MAPK in cells are important elements to
monitor when analysing the behaviour of the system, because of their
essential role in several signalling pathways and other cellular
processes.

An interesting aspect of this pathway is the presence of numerous
feedback mechanisms, which are carried out mainly by kinases and
phosphatases, regulating the amount of the involved proteins.

Two kinds of phosphatases are involved: first, the serine/threonine
protein phosphatase 2~(PP-2A), which is a ubiquitous enzyme able to
dephosphorylate a wide range of proteins among which there are the
components of the MAPK signalling cascade~(B-Raf, MEK and MAPK);
second, protein tyrosine phosphatases~(PTPs), which are also
involved in the dephosphorylation of MAPK.

In addition to its positive role carried out by activating B-Raf,
PKA is further involved in the regulation of the pathway, by
exercising both a positive and a negative regulatory role:
respectively, it phosphorylates PTPs~(in the phosphorylated form,
PTPs are less effective in dephosphorylating MAPK) and PDE4~(in the
phosphorylated form, PDE4 enhances the transformation of cAMP into
AMP, making it unavailable for activating PKA).

The relative strength of the different PKA enzymatic activities, and
particularly the opposed positive and negative feedbacks, is
therefore an interesting issue to be considered when studying PKA
regulatory role.

\subsection{A Bio-PEPA Model of the Network}
\label{sec:biomodel_biopepa}

In the schematic representation of the pathway in
Figure~\ref{fig:pathway} we have subdivided the system into three
different subnetworks:
\begin{itemize}
\item upstream pathway: G-protein mediated activation of AC
\item intermediate pathway: activation of cAMP-dependent PKA
\item downstream pathway: MAPK signalling cascade
\end{itemize}

The Bio-PEPA language allows the definition of compositional models
and, therefore, the above subdivision can be straightforwardly
employed in the model itself. This helps in improving the
readability of the model. The full Bio-PEPA model~(using the syntax
and the shorthand notation supported by the Bio-PEPA Eclipse
plug-in) is reported in Appendix~\ref{sec:appendix}.

Three locations are defined in the Bio-PEPA model: $cyto$,
$cyto\_mem$, and $extra$. Each of them represents one of the
involved biological compartments, and is assigned a size expressing
its volume~(3D compartments) or its surface area~(2D membranes).

Kinetic parameters and rate laws are specified. Some of the reaction
rates follow the mass-action law, while others are based on the
Michaelis-Menten kinetics. The quantitative data we adopt are those
defined in the original ODE model~\cite{sbml_NevesEtAl08}. The
reversible reactions are decomposed into two irreversible
(mass-action) reactions describing the forward and the backward
directions, respectively. The full set of parameters used is
reported in the supplementary material.

Each molecular species in the pathway is abstracted by a species
component describing the reactions in which it is involved and its
role in each of them~(i.e.~whether it is a reactant, a product, or
an enzyme). The Bio-PEPA model component describes how the species
components interact with each other, and defines the initial amounts
for each species.

There are two noteworthy differences with respect to the original
model. These are related to the conversions needed to switch to
molecular counts as units of measure for the species amount. In the
ODE model the species amounts are specified in terms of
concentrations~(numbers of molecules over volume/area), while in our
Bio-PEPA model they must be specified in terms of numbers of
molecules. Consequently, the rate laws must be rescaled accordingly
to take into account the sizes of the compartments in which the
species involved in the reactions are located.

In this definition of the model, compartment volumes and membrane
surface areas are assumed to be constant. This constraint is not
present in Bio-PEPA, which allows location sizes to be defined as
arbitrary functions~(e.g., they can change over time or depend on
the amount of given species). Here we do not consider generic
location sizes in order to be able to compare our results with the
original model. With this assumption, the system could be modelled
using languages which are not equipped with a notion of location
(provided reaction rates are carefully computed taking into account
volumes). However, using a language, such as Bio-PEPA, which is
equipped with a notion of location, has also the obvious advantage
that the explicit definition of information such as locations of
species, volumes and areas brings a greater clarity of the model.

\section{Model Analysis}
\label{sec:analysis}
\subsection{Validation}

In order to verify that the Bio-PEPA model is a faithful
representation of the original ODE model~\cite{sbml_NevesEtAl08}, we
compared the results we have obtained from the Bio-PEPA model with
the results obtained through the BioModels online simulator on the
original model.

\begin{figure}[htbp]
\centering
  \subfigure[Ratio of active MAPK]{\includegraphics[width=0.49\textwidth]{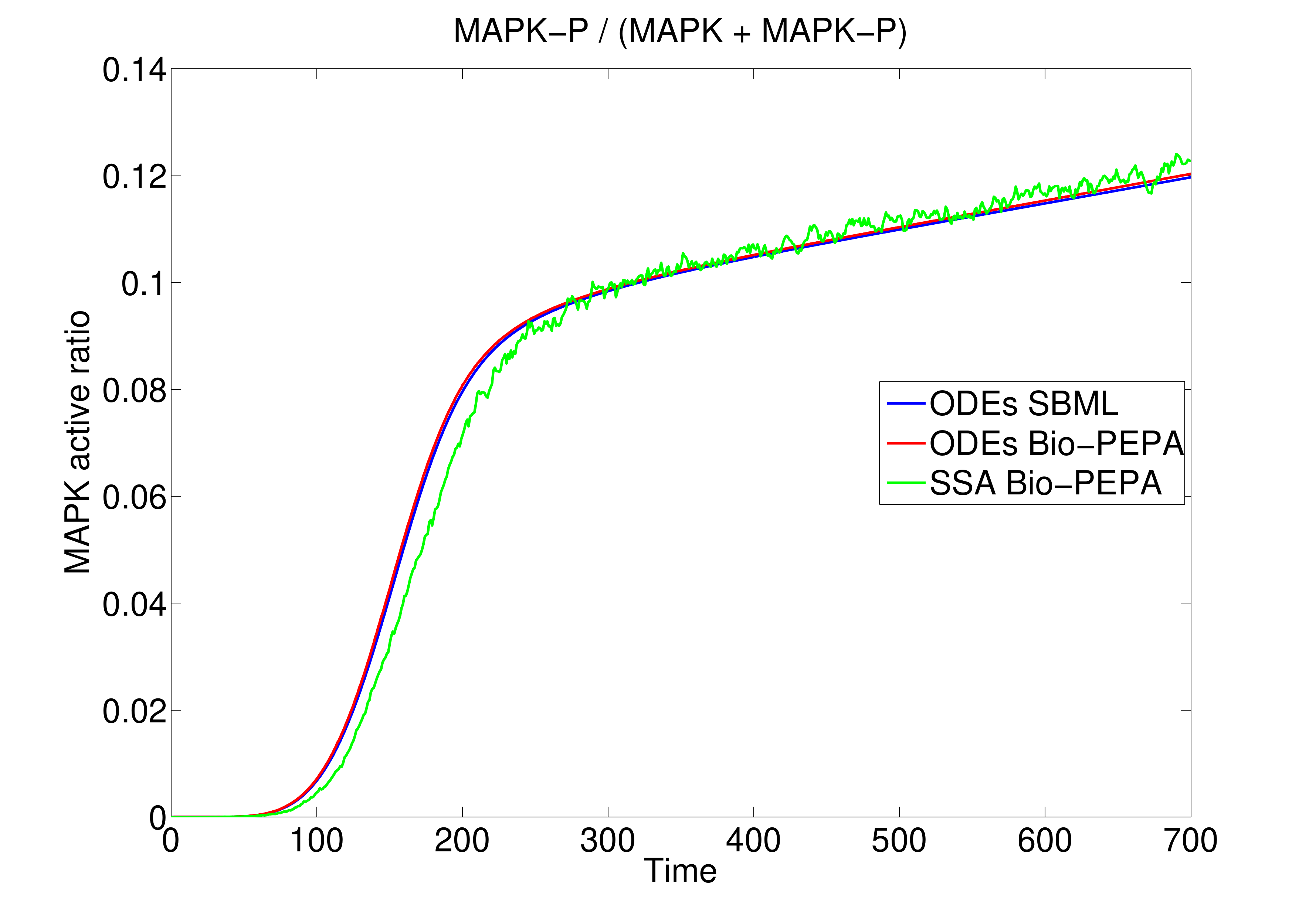}}
  \subfigure[Active AC, PKA and MAPK]{\includegraphics[width=0.49\textwidth]{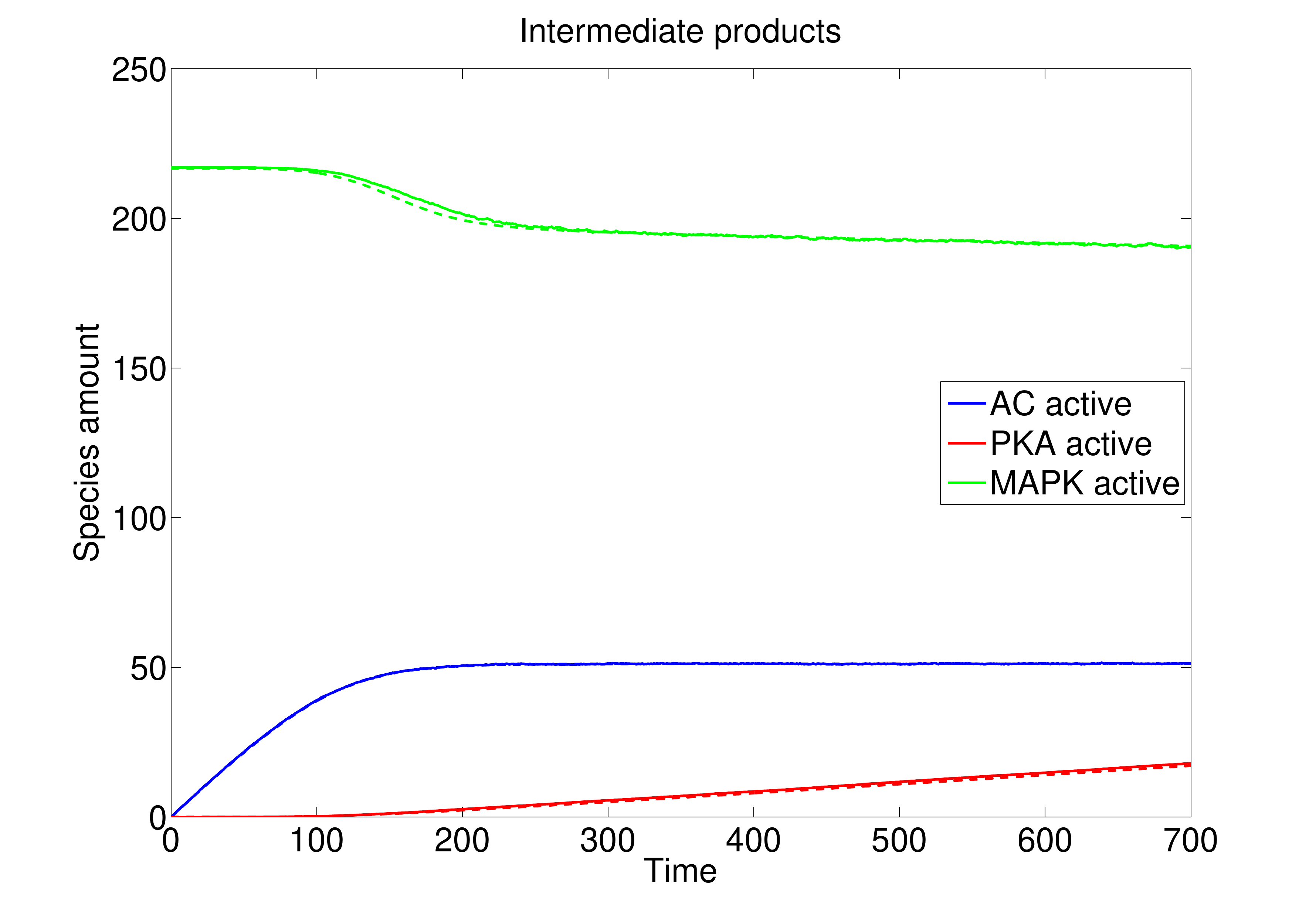}}
  \caption{Comparison between the simulation results obtained by the Bio-PEPA Eclipse Plug-in and the results of the original ODE model. SSA results are the average over 500 simulation runs.}
  \label{fig:comparison_sbml}
\end{figure}

Figure~\ref{fig:comparison_sbml}(a) reports the comparison of the
results for the ratio of active MAPK~(i.e.~phosphorylated over total
MAPK). We show both the time series obtained using one of the
stochastic simulators and the one obtained using one of the ODE
solvers available within the Bio-PEPA Eclipse Plug-in. We observe a
good agreement between the results from the Bio-PEPA model and the
ones from the original ODE model. In particular, the two ODE-based
analyses produce coincident results. The SSA plot, obtained as the
average over 500 individual stochastic simulation runs, is in good
agreement with the deterministic analysis and also shows the
variability of the system due to the presence of some stochastic
noise.

A similar agreement was also found for all the other molecular
species involved in the pathway. In
Figure~\ref{fig:comparison_sbml}(b) we report the results for three
of the key molecular species involved in the pathway~(AC, PKA, and
MAPK), each of which is the final product of one of the three
subnetworks. For each of these proteins, we compare the amount of
its active form in our SSA results~(solid lines) with the results
from the original model~(dashed lines).

Note that, in the SBML model, the amounts of species are given in
terms of numbers of molecules over compartment sizes and, hence, the
amounts must be rescaled by multiplying by the compartment sizes in
order to be comparable with our results, which are expressed in
terms of molecular counts. AC is located on the cell membrane and,
therefore, it must be multiplied by the area of the membrane
($0.2~\mu m^2$); in the case of PKA and MAPK the amounts result the
same, since the volume of the cytoplasm is $1~\mu m^3$.

\subsection{Experimentation}

In this section we report some of the computational experiments we
have performed on the model. First, we have studied the effect, on
the final product of the pathway~(MAPK-P, i.e.~phosphorylated MAPK),
of changes in the network structure: we have removed some of the
feedback loops, in particular the ones involving PKA, which is one
of the proteins mostly responsible for both positive and negative
regulation of various pathway components.

\begin{figure}[htbp]
\centering
  \subfigure[Reduction of some reactions rates]{\includegraphics[width=0.49\textwidth]{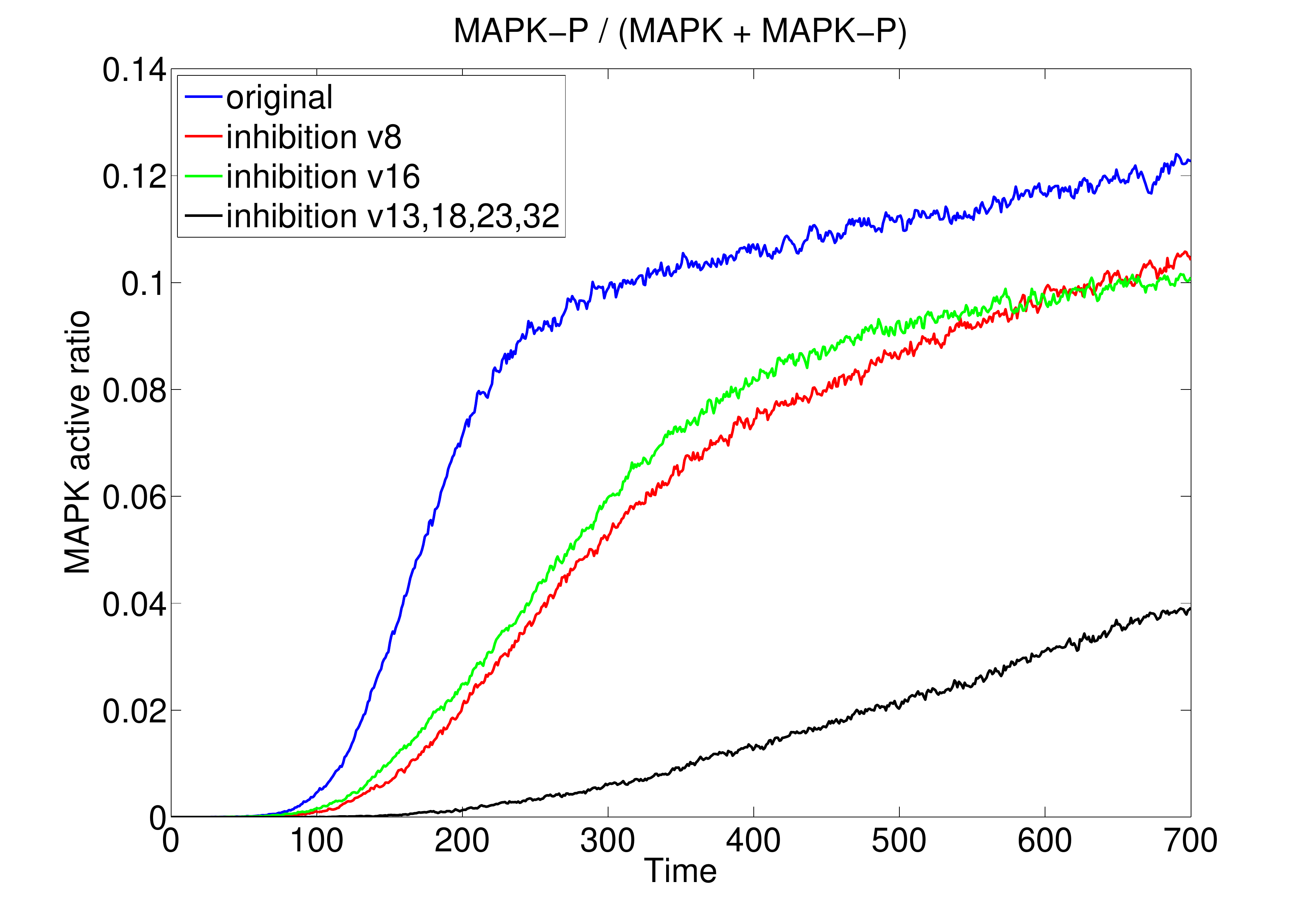}}
  \subfigure[Inhibition/strengthening of PKA feedbacks]{\includegraphics[width=0.49\textwidth]{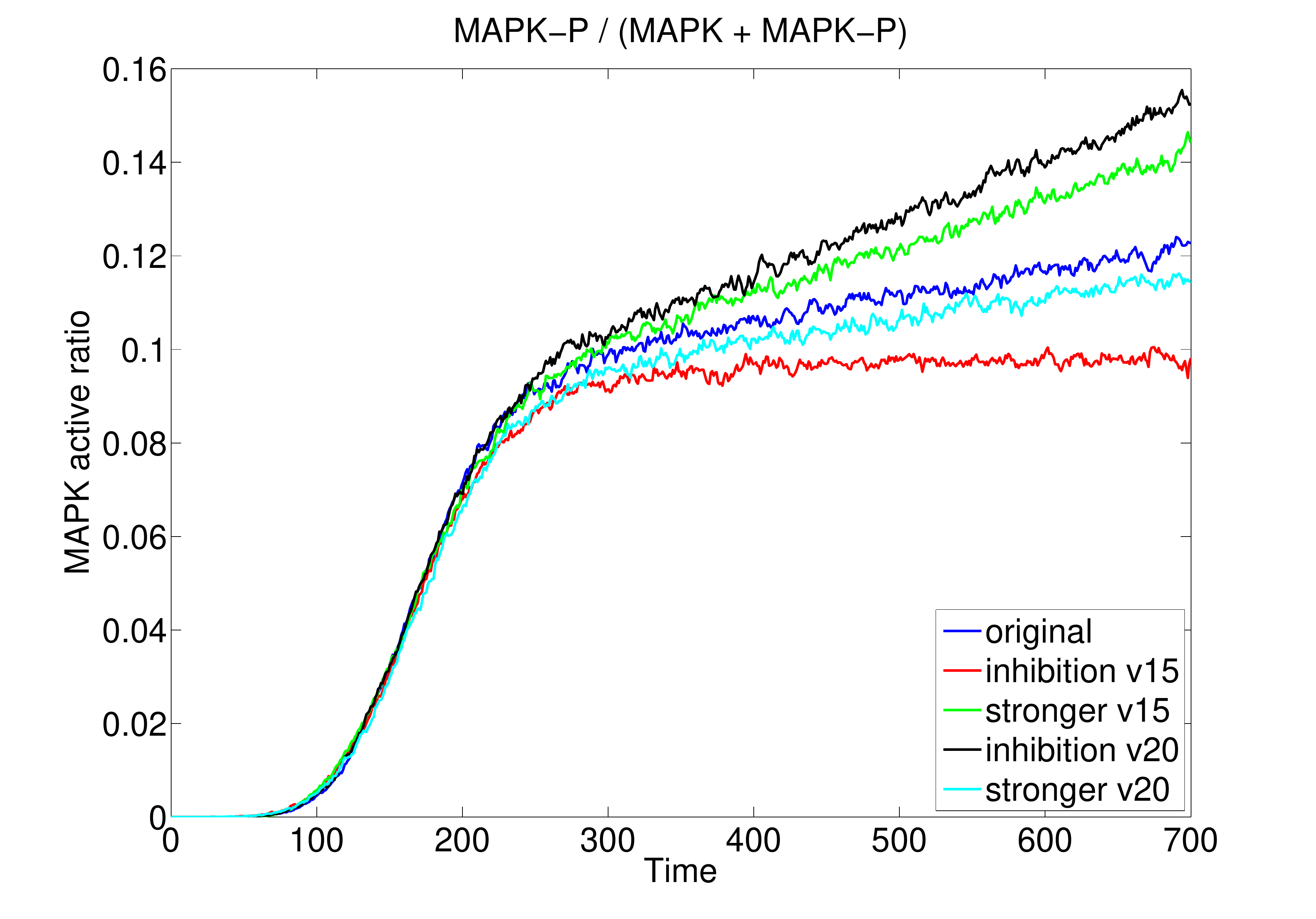}}
  \caption{Experimenting on reaction rates and feedback loops~(each line is the average
behaviour over 500 independent stochastic simulation runs).}
  \label{fig:inhibition}
\end{figure}

Figure~\ref{fig:inhibition} shows the effect of inhibiting some
important steps on the behaviour of the system. In
Figure~\ref{fig:inhibition}(a) we compare the behaviour of the
original system~(blue line) with the behaviour of three modified
systems: one with weaker enzymatic activity of PKA~(10-fold increase
of $K_m\_v08$) in the phosphorylation of B-Raf~(red line), one with
slower activation of AC~(10-fold decrease of $K_f\_AC\_activation$
in the binding reaction between AC and $G_{\alpha s}:GTP$)~(green
line), and one with slower cAMP/R2C2 binding~(10-fold decrease of
$K_f\_v13$, $K_f\_v18$, $K_f\_v23$, $K_f\_v32$)~(black line). Each
of these modified systems corresponds to slowing down one reaction
rate in each of the subnetworks; Figure~\ref{fig:inhibition}(a)
shows that a reduction in the rate of cAMP binding has the strongest
effect: an expected result, since the cAMP binding reaction must
occur four times for a single molecule of PKA to become active.

In Figure~\ref{fig:inhibition}(b), instead, we experiment with the
model by changing the parameters relative to the enzymatic activity
of PKA and studying their effect on the system behaviour. As
described in Section~\ref{sec:biomodel}, PKA takes part in a
positive feedback loop by phosphorylating PTPs, and in a negative
feedback loop by phosphorylating PDE4.

The results shown in Figure~\ref{fig:inhibition}(b) confirm the
presence of these feedback loops: a weaker enzymatic activity of PKA
(100-fold increase of $K_m\_v15$) in the phosphorylation of PTP~(red
line) leads to a decrease of MAPK-P~(due to the lower presence of
phosphorylated PTP, which is less effective than its
unphosphorylated form in dephosphorylating MAPK), while a stronger
enzymatic activity of PKA~(100-fold decrease of $K_m\_v15$)~(green
line) leads to an increase of MAPK-P. The opposite occurs for the
phosphorylation of PDE4: a weaker enzymatic activity of PKA
(100-fold increase of $K_m\_v20$) in the phosphorylation of PDE4
(black line) leads to an increase of MAPK-P~(due to the lower
presence of phosphorylated PDE4, which is more effective than its
unphosphorylated form in transforming cAMP into AMP), while a
stronger enzymatic activity of PKA~(100-fold decrease of $K_m\_v20$)
(cyan line) leads to a decrease~(though smaller) of MAPK-P.

\section{Related work}
\label{sec:relatedworks} Several languages have been defined to
model biological systems with compartments and
membranes~\cite{paun-rozenberg02,cardelli04,regevEtAl04,priami-quaglia05a,Cavaliere06,versari08,Laneve08}.
In the following we report a brief overview of the most well-known
languages and the associated analysis tools.

\emph{Membrane systems} or \emph{P
systems}~\cite{paun-rozenberg02,Ciobanu06} are computational models that are based on a notion of membrane structure. In particular, multisets of objects are enclosed in a nested hierarchy of membranes and their behaviour is described by local rewriting rules.

The analysis of P systems is supported by the \emph{P System
Modelling Framework}~\cite{frameworkHP}. This framework provides
stochastic simulation based on the multi-compartmental Gillespie
algorithm~\cite{romero06} and offers the mapping to
PRISM~\cite{hintonEtAl06,prism_site} for model checking.
Furthermore, translations to \emph{SBML}~\cite{sbml} and
\emph{CellDesigner}~\cite{celldesigner} have been implemented in
order to export P system  models into other formats for further
kinds of analysis and for graphical representation.

The variant of membrane systems with peripheral and integral
membrane proteins defined by Cavaliere and
Sedwards~\cite{Cavaliere06}  is at the basis  of
\emph{Cyto-Sim}~\cite{Sedwards07,cytosimHP}. The analysis is limited
to stochastic simulation by Gillespie's
algorithm~\cite{gillespie77}, but it is possible to export the
corresponding ODE model into MATLAB format~\cite{MATLAB} for further
analyses. In addition to mass-action kinetics, the simulator
supports chemical reactions with arbitrary kinetic laws based on
functions of the reactants.

One of the first process calculi with an explicit notion of
compartments is \emph{BioAmbients} calculus~\cite{regevEtAl04}. A
BioAmbient system is seen as a hierarchy of nested \emph{ambients},
generally abstracting compartments,  containing communicating
processes whose actions specify the evolution of the system. Various
kinds of action involving compartments can be easily represented,
such as transport of small molecules across compartments.
BioAmbients models are supported by two tools:
\emph{Bio-SPI}~\cite{silverman}, a stochastic simulator based on
Gillespie's algorithm and originally implemented for models in
stochastic $\pi$-calculus, and the  \emph{BioAmbient machine
(BAM)}~\cite{vigliotti08,BAMHP}, a  more efficient and user-friendly
simulator written in Java. BAM is based on the stochastic version of
BioAmbients, defined by Vigliotti and Harrison~\cite{vigliotti06}
and can be used to simulate bio-regulatory pathways as well as
membrane interaction.  BAM automatically produces both a simulation
graph and a debugging trace of the program. Furthermore, it offers a
graphical user interface for entering program code and visualising
simulation results.

A more active role for membranes can be described using \emph{Brane calculi}~\cite{cardelli04}.
In Brane calculi, a system is represented as a set
of nested membranes, and a membrane is represented as a set of
actions. Membranes can move,
merge, split, enter into and exit from other membranes.

The bio$\kappa$-\textit{calculus}~\cite{Laneve08} is a language for
describing proteins and cells based on Brane calculi and
$\kappa$-calculus~\cite{danos-laneve04}. The atomic elements of the
calculus are  \textit{proteins} and two constructors for
representing \textit{solutions} and \textit{cells}. Proteins are
characterised by sites, that can be in different states:
\textit{bound} to another site of a protein, \textit{visible} (i.e\
not connected to other sites) and \textit{hidden} (i.e \ not
connected to other sites but not available for other interactions).
Cells abstract compartments and consist of a membrane and a
cytoplasm.  Reactions can be interactions between two proteins or
between two membranes.

In \textit{Beta-binders}~\cite{priami-quaglia05a} and the associated
\emph{BlenX} language~\cite{romanel08}, systems are modelled as a
composition of boxes representing biological entities. In the
original version of the language compartments are not defined
explicitly, but a virtual form of nesting is rendered by appropriate
typing for sites. Explicit static compartments and transport of
biological across them have been added to Beta-binders
in~\cite{guerriero-priami-romanel07}. The analysis of BlenX models
is possible in the  \emph{Beta Workbench (BetaWB)}~\cite{betawbHP}.
The BetaWB is composed of a stochastic simulator based on an
efficient variant of  Gillespie's algorithm, the BetaWB designer, a
graphical editor for developing models and the BetaWB plotter, a
tool to analyse the results of a stochastic simulation run.
Furthermore, BlenX models can be exported in SBML.

Finally, there are some variants of the stochastic $\pi$-calculus
enriched with the notion of locations (see~\cite{versari08,john08},
for instance). S$\pi$@ calculus~\cite{versari08} extends the
stochastic $\pi$-calculus syntax with the explicit addition of
compartments. This language handles varying volumes and dynamic
compartments by defining the compartment volume as the sum of the
volumes occupied by all the molecules it contains.

Another extension of the $\pi$-calculus with compartments is
\emph{SpacePi}~\cite{john08}. Pi processes are embedded into a
vector space and move individually. Only processes that are
sufficiently close can communicate. The operational semantics of
SpacePi defines the interplay between movement, communication, and
time-triggered events.

All the languages listed above differ from Bio-PEPA in various aspects: they are based on different levels of abstraction or focus on dynamical compartments or handle volumes in a distinct way. In Bio-PEPA, locations have a fixed structure and compartments are essentially containers for biological species: interactions between molecules in different
compartments is allowed, but the main evolution is given by interactions of
molecular species within compartments.
This concept of location is that usually considered in biochemical networks
in databases and in the literature.
For these systems Bio-PEPA offers a direct and formal representation, allowing an
intuitive representation of both intra-compartment and
inter-compartment reactions.

Concerning the Bio-PEPA Eclipse Plug-in, a prominent feature of the
tool is its support of both stochastic simulators and ODE numerical
solvers. The ability to utilise different analysis techniques is
very important; this can aid in understanding different behavioural
aspects of the system and discovering possible errors due to the use
of a specific solver/simulator~\cite{calderEtAl06}. Furthermore, the
user can select the most appropriate analysis method based on the
model under study. Most of the other tools support stochastic
simulation by Gillespie's algorithm only and in a few cases mappings
to other languages or tools, such as SBML or PRISM.

\section{Conclusions}
\label{sec:conclusions}

Interactions of molecular species located in compartments with
different sizes occur typically in biological systems. In signalling
pathways, for instance, a biochemical signal is transferred from the
extra-cellular environment into the cell through the intervention of
membrane receptors and other membrane-bound proteins.

In this work we have presented a Bio-PEPA compartmental model of the cAMP/PKA/MAPK pathway. Our main aim was to demonstrate, on a real case study, the effectiveness and the correctness of the representation of multi-compartment models in Bio-PEPA.

The Bio-PEPA language allows us to explicitly define locations representing compartments and membranes, to express the location of each species and to indicate the relative position of a location with respect to the others. Furthermore, Bio-PEPA correctly deals with the location  sizes~(volumes and surface areas). It is generally useful to collect these kinds of information, in order to have a better view of the system under consideration. Moreover, the information about the size of locations is necessary to compare the results obtained in Bio-PEPA with the ones in the literature and experiments, when given in concentrations instead of number of molecules. Another benefit is although the system in question can be written as a set of ODEs, high-level languages like Bio-PEPA can offer a cleaner view of the model through a more restricted~(but focussed) number of operators. Through these operators the modeller must explicitly state any behaviours he or she wishes the components of the system to exhibit. A restricted set of operators also allows the tool to perform more extensive checking over any legal Bio-PEPA model, something not possible if manually coding a set of ODEs in a general tool such as MATLAB.

In order to be able to validate our results, we developed our
Bio-PEPA model as a faithful representation of an existing ODE
model.

The analysis of the pathway was performed using the stochastic simulator available within the Bio-PEPA Eclipse Plug-in, which is able to handle multi-compartmental models. The comparison of our results with the ones obtained from the original ODE model demonstrate the correctness of our approach. We also perform a number of simulation experiments in order to investigate specific properties of the system such as its response to the inhibition of feedback mechanisms and to the variation of key molecular species.

In this model we have found it convenient to describe the behaviour
of the pathway in terms of three functional elements, a feature
supported by the tool. Currently such modularisation is supported
only at the level of model definition as a convenience for the
modeller. However it is an interesting area for future work to
explore the extent to which such structuring might be exploited
during model analysis.

While the locations extension for Bio-PEPA does not restrict species interaction by location, future improvements within the tool will flag such reactions. Once implemented this will allow the modeller to clearly see which reactions involve species not existing in adjacent locations. Other future work in the Bio-PEPA Eclipse Plug-in includes improved support for locations as well as the implementation for automatic import/export of existing specification languages supporting multi-compartment models~(e.g.~\cite{sbml,sbgn,guerriero-heath-priami07}).

Finally, we would like to extend the analysis of this model~(and of
multi-compartmental models in general) by applying some of the
CTMC-based analysis methods~(e.g.~model-checking) which are
supported by the Bio-PEPA language.

\subsection*{Acknowledgements}
The authors thank Jane Hillston for her helpful comments. At the
time of writing this paper, Federica Ciocchetta was a research
fellow at the University of Edinburgh, supported by the EPSRC grant
EP/C543696/1 ``Process Algebra Approaches to Collective Dynamics''.
Adam Duguid is supported by the EPSRC Doctoral Training Grant
EP/P501407/1. Maria Luisa Guerriero is supported by the EPSRC grant
EP/E031439/1 ``Stochastic Process Algebra for Biochemical Signalling
Pathway Analysis''.

\bibliographystyle{eptcs} 
\bibliography{biblio}
\appendix
\section{Bio-PEPA Model}\label{sec:appendix}
\setlength{\tabcolsep}{5pt}

In this appendix we report the full Bio-PEPA model, specified in the tool syntax,  corresponding to the cAMP/PKA/B-Raf/MAPK1,2 pathway. The abbreviations described in the paper are used.
The measure unit for the volume size is $\mu \,m^3$ and the unit for the area is $\mu \,m^2$. The units of parameters are defined accordingly.

In the definition of parameters, the constants $omega\_cyto$ and $omega\_extra$ are the scaling factors used for converting parameters involving concentrations~(expressed in $\mu \,M$) in the correct unit in terms of number of molecules. $omega\_cyto$
is for parameters of reactions in the cytoplasm whereas $omega\_extra$ is for parameters of reactions in the extracellular space. $omega\_cyto$ is defined as $10^{-6} \cdot N_A \cdot size(cyto)$, where $10^{-6}$ is to convert $\mu \,M$ into $M$, $N_A=6.022 \cdot 10^{23}$ is the Avogadro number~(i.e.\ the number of molecules in a mole of substance) and $ size(cyto)$ is the volume size of the cytoplasm. Similarly, $omega\_extra = 10^{-6} \cdot N_A \cdot size(extra)$, where $size(extra)$ is the volume size for the extracellular space.

\begin{small}
\begin{longtable}[l]{l l l}
location cyto in cyto\_mem & : size = 1, & kind = compartment;\\
location extra & : size = 0.111, & kind = compartment;\\
location cyto\_mem  in extra & : size = 0.2, & kind = membrane;
\end{longtable}

\begin{longtable}[l]{l l l | l l l}
omega\_cyto & = & 602; & omega\_extra & = & 66.82;\\
&&&\\
kcat\_pde4\_p\_pde4\_p & = & 20; & kcat\_PPase\_Raf & = & 5;\\
kcat\_PDE4\_PDE4 & = & 8; & kcat\_MEK\_activates\_MAPK & = & 0.15;\\
kcat\_PKA\_activates\_Raf & = & 10; & kcat\_AC\_active\_AC\_active & = &  8.5;\\
kcat\_highKM\_PDE & = & 8; & kcat\_PKA\_P\_PTP & = & 0.2;\\
kcat\_AC\_basal\_AC\_basal & = & 0.2; & kcat\_grk\_GRK  & = & 0.104;\\
kcat\_PKA\_P\_PDE & = & 10; & kcat\_Raf\_activates\_MEK & = & 0.105;\\
kcat\_PTP\_PKA & = & 0.1; & kcat\_PTP & = & 1.06;\\
kcat\_PPase\_MAPK & = & 0.636; & kcat\_pp2a\_4\_pp2a\_4 & = & 5;\\
kcat\_pp\_ptp\_pp\_ptp & = & 5; & kcat\_GRK\_bg\_GRK\_bg & = & 1.34;\\
kcat\_PPase\_mek & = & 5;\\
&&&\\
Kf\_activate\_Gs & = & 0.025; & Kr\_activate\_Gs & = & 0;\\
Km\_pde4\_p & = & 1.3; & Km\_v3 & = & 15.7;\\
Kf\_v4 & = & 1; & Kr\_v4 & = & 0.2;\\
Km\_PDE4 & = & 1.3; & Kf\_bg\_binds\_GRK & = & 1;\\
Kr\_bg\_binds\_GRK & = & 0.5; & Km\_v07 & = & 0.046;\\
Km\_v08 & = & 0.5; & Km\_AC\_active & = & 32;\\
Kf\_GTPase & = & 0.067; & Kr\_GTPase & = & 0;\\
Kf\_trimer & = & 6; & Kr\_trimer & = & 0;\\
Kf\_G\_binds\_iso\_BAR & = & 10; & Kr\_G\_binds\_iso\_BAR & = & 0.1;\\
Kf\_v13 & = & 8.35; & Kr\_v13 & = & 0.017;\\
Km\_v14 & = & 15; & Km\_v15 & = & 0.1;\\
Kf\_AC\_activation & = & 500; & Kr\_AC\_activation & = & 1;\\
Km\_AC\_basal & = & 1030; & Kf\_v18 & = & 0.006;\\
Kr\_v18 & = & 0.00028; & Km\_grk & = & 15;\\
Km\_v20 & = & 0.5; & Km\_v21 & = & 0.159;\\
Km\_v22 & = & 9; & Kf\_v23 & = & 0.006;\\
Kr\_v23 & = & 2.8e-4; & Km\_v24 & = & 0.46;\\
Kf\_v25 & = & 1; & Kr\_v25 & = & 0.062;\\
Km\_v26 & = & 0.77; & Km\_pp2a\_4 & = & 8;\\
Km\_v28 & = & 6; & Km\_GRK\_bg & = & 4;\\
Kf\_G\_binds\_BAR & = & 0.3; & Kr\_G\_binds\_BAR & = & 0.1;\\
Km\_v31 & = & 15.7; & Kf\_v32 & = & 8.350;\\
Kr\_v32 & = & 0.017; &
\end{longtable}

\begin{longtable}[l]{l l}
kineticLawOf v01f &: fMA(Kf\_activate\_Gs);\\
kineticLawOf v01b &: fMA(Kr\_activate\_Gs /(omega\_cyto * omega\_cyto));\\
kineticLawOf v02 &: fMM(kcat\_pde4\_p\_pde4\_p, Km\_pde4\_p * omega\_cyto);\\
kineticLawOf v03 &: fMM(kcat\_PPase\_Raf, Km\_v3 * omega\_cyto);\\
kineticLawOf v04f &: fMA(Kf\_v4 /omega\_extra);\\
kineticLawOf v04b &: fMA(Kr\_v4);\\
kineticLawOf v05 &: fMM(kcat\_PDE4\_PDE4, Km\_PDE4 * omega\_cyto);\\
kineticLawOf v06f &: fMA(Kf\_bg\_binds\_GRK / omega\_cyto);\\
kineticLawOf v06b &: fMA(Kr\_bg\_binds\_GRK);\\
kineticLawOf v07 &: fMM(kcat\_MEK\_activates\_MAPK, Km\_v07 * omega\_cyto);\\
kineticLawOf v08 &: fMM(kcat\_PKA\_activates\_Raf, Km\_v08 * omega\_cyto);\\
kineticLawOf v09 &: fMM(kcat\_AC\_active\_AC\_active, Km\_AC\_active * omega\_cyto);\\
kineticLawOf v10f &: fMA(Kf\_GTPase);\\
kineticLawOf v10b &: fMA(Kr\_GTPase);\\
kineticLawOf v11f &: fMA(Kf\_trimer / omega\_cyto);\\
kineticLawOf v11b &: fMA(Kr\_trimer);\\
kineticLawOf v12f &: fMA(Kf\_G\_binds\_iso\_BAR/omega\_cyto);\\
kineticLawOf v12b &: fMA(Kr\_G\_binds\_iso\_BAR);\\
kineticLawOf v13f &: fMA(Kf\_v13 / omega\_cyto);\\
kineticLawOf v13b &: fMA(Kr\_v13);\\
kineticLawOf v14 &: fMM(kcat\_highKM\_PDE, Km\_v14 * omega\_cyto) ;\\
kineticLawOf v15 &: fMM(kcat\_PKA\_P\_PTP, Km\_v15 * omega\_cyto);\\
kineticLawOf v16f &: fMA(Kf\_AC\_activation/omega\_cyto);\\
kineticLawOf v16b &: fMA(Kr\_AC\_activation);\\
kineticLawOf v17 &: fMM(kcat\_AC\_basal\_AC\_basal, Km\_AC\_basal*omega\_cyto);\\
kineticLawOf v18f &: fMA(Kf\_v18 /omega\_cyto);\\
kineticLawOf v18b &: fMA(Kr\_v18);\\
kineticLawOf v19 &: fMM(kcat\_grk\_GRK * cyto\_mem / omega\_cyto, Km\_grk * cyto\_mem);\\
kineticLawOf v20 &: fMM(kcat\_PKA\_P\_PDE, Km\_v20*omega\_cyto);\\
kineticLawOf v21 &: fMM(kcat\_Raf\_activates\_MEK, Km\_v21 * omega\_cyto);\\
kineticLawOf v22 &: fMM(kcat\_PTP\_PKA, Km\_v22 * omega\_cyto);\\
kineticLawOf v23f &: fMA(Kf\_v23 /omega\_cyto) ;\\
kineticLawOf v23b &: fMA(Kr\_v23);\\
kineticLawOf v24 &: fMM(kcat\_PTP, Km\_v24 *omega\_cyto);\\
kineticLawOf v25f &: fMA(Kf\_v25/omega\_extra);\\
kineticLawOf v25b &: fMA(Kr\_v25);\\
kineticLawOf v26 &: fMM(kcat\_PPase\_MAPK, Km\_v26 * omega\_cyto);\\
kineticLawOf v27 &: fMM(kcat\_pp2a\_4\_pp2a\_4, Km\_pp2a\_4 *omega\_cyto);\\
kineticLawOf v28 &: fMM(kcat\_pp\_ptp\_pp\_ptp, Km\_v28 * omega\_cyto);\\
kineticLawOf v29 &: fMM(kcat\_GRK\_bg\_GRK\_bg * cyto\_mem /omega\_cyto, Km\_GRK\_bg * cyto\_mem);\\
kineticLawOf v30f &: fMA(Kf\_G\_binds\_BAR/omega\_cyto);\\
kineticLawOf v30b &: fMA(Kr\_G\_binds\_BAR);\\
kineticLawOf v31 &: fMM(kcat\_PPase\_mek, Km\_v31 * omega\_cyto);\\
kineticLawOf v32f &: fMA(Kf\_v32 /omega\_cyto);\\
kineticLawOf v32b &: fMA(Kr\_v32);
\end{longtable}

\begin{longtable}[l]{l l l}
AC\_active & = & v16f $>>$ AC\_active@cyto\_mem + v16b $<<$ AC\_active@cyto\_mem +\\
&& v09 (+) AC\_active@cyto\_mem;\\
G\_GDP & = & v10f $>>$ G\_GDP@cyto + v10b $<<$ G\_GDP@cyto +\\
&& v11f $<<$ G\_GDP@cyto + v11b $>>$ G\_GDP@cyto;\\
G\_protein & = & v11f $>>$ G\_protein@cyto + v11b $<<$ G\_protein@cyto +\\
&& v12f $<<$ G\_protein@cyto + v12b $>>$ G\_protein@cyto +\\
&& v30f $<<$ G\_protein@cyto + v30b $>>$ G\_protein@cyto;\\
G\_a\_s & = & v01f $>>$ G\_a\_s@cyto + v01b $<<$ G\_a\_s@cyto + v10f $<<$ G\_a\_s@cyto +\\
&& v10b $>>$ G\_a\_s@cyto + v16f $<<$ G\_a\_s@cyto + v16b $>>$ G\_a\_s@cyto;\\
GRK\_bg & = & v06f $>>$ GRK\_bg@cyto + v06b $<<$ GRK\_bg@cyto +\\
&& v29 (+) GRK\_bg@cyto;\\
iso\_BAR\_p & = & v19 $>>$ iso\_BAR\_p@cyto\_mem + v29 $>>$ iso\_BAR\_p@cyto\_mem;\\
PDE4 & = & v27 $>>$ PDE4@cyto + v20 $<<$ PDE4@cyto + v05 (+) PDE4@cyto;\\
ATP & = & v09 $<<$ ATP@cyto + v17 $<<$ ATP@cyto;\\
R2C2 & = & v18f $<<$ R2C2@cyto + v18b $>>$ R2C2@cyto;\\
PP\_PDE & = & v27 (+) PP\_PDE@cyto;\\
BAR & = & v04f $<<$ BAR@cyto\_mem + v04b $>>$ BAR@cyto\_mem +\\
&& v30f $<<$ BAR@cyto\_mem + v30b $>>$ BAR@cyto\_mem;\\
BAR\_G & = & v30f $>>$ BAR\_G@cyto\_mem + v30b $<<$ BAR\_G@cyto\_mem +\\
&& v25f $<<$ BAR\_G@cyto\_mem + v25b $>>$ BAR\_G@cyto\_mem;\\
iso & = & v04f $<<$ iso@extra + v04b $>>$ iso@extra + v25f $<<$ iso@extra +\\
&& v25b $>>$ iso@extra;\\
iso\_BAR & = & v01f $>>$ iso\_BAR@cyto\_mem + v01b $<<$ iso\_BAR@cyto\_mem +\\
&& v04f $>>$ iso\_BAR@cyto\_mem + v04b $<<$ iso\_BAR@cyto\_mem +\\
&& v12f $<<$ iso\_BAR@cyto\_mem + v12b $>>$ iso\_BAR@cyto\_mem +\\
&& v19 $<<$ iso\_BAR@cyto\_mem + v29 $<<$ iso\_BAR@cyto\_mem;\\
MAPK\_active & = & v07 $>>$ MAPK\_active@cyto + v22 $<<$ MAPK\_active@cyto +\\
&& v24 $<<$ MAPK\_active@cyto + v26 $<<$ MAPK\_active@cyto;\\
MEK & = & v31 $>>$ MEK@cyto + v21 $<<$ MEK@cyto;\\
MEK\_active & = & v21 $>>$ MEK\_active@cyto + v31 $<<$ MEK\_active@cyto +\\
&& v07 (+) MEK\_active@cyto;\\
B\_Raf\_active & = & v08 $>>$ B\_Raf\_active@cyto + v03 $<<$ B\_Raf\_active@cyto +\\
&& v21 (+) B\_Raf\_active@cyto;\\
bg & = & v01f $>>$ bg@cyto + v01b $<<$ bg@cyto + v06f $<<$ bg@cyto +\\
&& v06b $>>$ bg@cyto + v11f $<<$ bg@cyto + v11b $>>$ bg@cyto;\\
B\_Raf & = & v03 $>>$ B\_Raf@cyto + v08 $<<$ B\_Raf@cyto;\\
PKA & = & v13f $>>$ PKA@cyto + v13b $<<$ PKA@cyto + v08 (+) PKA@cyto +\\
&& v15 (+) PKA@cyto + v20 (+) PKA@cyto;\\
AC & = & v16f $<<$ AC@cyto\_mem + v16b $>>$ AC@cyto\_mem +\\
&& v17 (+) AC@cyto\_mem;\\
AMP & = & v02 $>>$ AMP@cyto + v05 $>>$ AMP@cyto + v14 $>>$ AMP@cyto;\\
GRK & = & v06f $<<$ GRK@cyto + v06b $>>$ GRK@cyto + v19 (+) GRK@cyto;\\
PP2A & = & v03 (+) PP2A@cyto + v26 (+) PP2A@cyto + v31 (+) PP2A@cyto;\\
MAPK & = & v22 $>>$ MAPK@cyto + v24 $>>$ MAPK@cyto + v26 $>>$ MAPK@cyto +\\
&& v07 $<<$ MAPK@cyto;\\
PTP & = & v28 $>>$ PTP@cyto + v15 $<<$ PTP@cyto + v24 (+) PTP@cyto;\\
PTP\_PKA & = & v15 $>>$ PTP\_PKA@cyto + v28 $<<$ PTP\_PKA@cyto +\\
&& v22 (+) PTP\_PKA@cyto;\\
c\_R2C2 & = & v18f $>>$ c\_R2C2@cyto + v18b $<<$ c\_R2C2@cyto +\\
&& v23f $<<$ c\_R2C2@cyto + v23b $>>$ c\_R2C2@cyto;\\
c2\_R2C2 & = & v23f $>>$ c2\_R2C2@cyto + v23b $<<$ c2\_R2C2@cyto +\\
&& v32f $<<$ c2\_R2C2@cyto + v32b $>>$ c2\_R2C2@cyto;\\
c3\_R2C2 & = & v32f $>>$ c3\_R2C2@cyto + v32b $<<$ c3\_R2C2@cyto +\\
&& v13f $<<$ c3\_R2C2@cyto +  v13b $>>$ c3\_R2C2@cyto;\\
iso\_BAR\_G & = & v12f $>>$ iso\_BAR\_G@cyto\_mem + v12b $<<$ iso\_BAR\_G@cyto\_mem +\\
&& v25f $>>$ iso\_BAR\_G@cyto\_mem + v25b $<<$ iso\_BAR\_G@cyto\_mem +\\
&& v01f $<<$ iso\_BAR\_G@cyto\_mem +v01b $>>$ iso\_BAR\_G@cyto\_mem;\\
PDE\_high\_km & = & v14 (+) PDE\_high\_km@cyto;\\
cAMP & = & v09 $>>$ cAMP@cyto + v17 $>>$ cAMP@cyto + v02 $<<$ cAMP@cyto +\\
&& v05 $<<$ cAMP@cyto + v13f $<<$ cAMP@cyto + v13b $>>$ cAMP@cyto +\\
&& v14 $<<$ cAMP@cyto + v18f $<<$ cAMP@cyto + v18b $>>$ cAMP@cyto +\\
&& v23f $<<$ cAMP@cyto + v23b $>>$ cAMP@cyto + v32f $<<$ cAMP@cyto +\\
&& v32b $>>$ cAMP@cyto;\\
PTP\_PP & = & v28 (+) PTP\_PP@cyto;\\
PDE4\_P & = & v20 $>>$ PDE4\_P@cyto + v27 $<<$ PDE4\_P@cyto + v02 (+) PDE4\_P;\\
\end{longtable}

\begin{longtable}[l]{l l l}
\multicolumn{3}{l}{MAPK\_active\_fraction  = MAPK\_active@cyto / (MAPK@cyto + MAPK\_active@cyto);}\\
\\
MAPK\_Pathway &::=& B\_Raf@cyto[120] $<*>$ B\_Raf\_active@cyto[0] $<*>$ MEK@cyto[108] $<*>$\\
&& MEK\_active@cyto[0] $<*>$ MAPK@cyto[217] $<*>$\\
&& MAPK\_active@cyto[0] $<*>$ PP2A@cyto[60] $<*>$  PTP@cyto[120] $<*>$\\
&& PTP\_PKA@cyto[0] $<*>$ PTP\_PP@cyto[60];\\
\\
cAMP\_Pathway &::=& cAMP@cyto[0] $<*>$ ATP@cyto[3010000] $<*>$ PKA@cyto[0] $<*>$\\
&& PDE4\_P@cyto[0] $<*>$ PDE4@cyto[241] $<*>$ PP\_PDE@cyto[120] $<*>$\\
&& PDE\_high\_km@cyto[301] $<*>$ AMP@cyto[3010000] $<*>$\\
&& R2C2@cyto[120]  $<*>$ c\_R2C2@cyto[0] $<*>$ c2\_R2C2@cyto[0] $<*>$\\
&& c3\_R2C2@cyto[0];\\
\\
G\_Pathway &::=& iso@extra[668] $<*>$ BAR@cyto\_mem[19] $<*>$ G\_protein@cyto[2167] $<*>$\\
&& iso\_BAR@cyto\_mem[0] $<*>$ iso\_BAR\_p@cyto\_mem[0] $<*>$\\
&& iso\_BAR\_G@cyto\_mem[0] $<*>$ G\_a\_s@cyto[0] $<*>$ AC@cyto\_mem[60] $<*>$\\
&& AC\_active@cyto\_mem[0] $<*>$ G\_GDP@cyto[0] $<*>$ GRK\_bg@cyto[0] $<*>$\\
&& BAR\_G@cyto\_mem[0] $<*>$ bg@cyto[0] $<*>$ GRK@cyto[1];\\
\\
\multicolumn{3}{l}{G\_Pathway $<$v09, v17$>$ cAMP\_Pathway $<$v08, v15$>$ MAPK\_Pathway}
\end{longtable}
\end{small}

\end{document}